\documentclass[12pt]{article}
\usepackage{graphicx}

\begin{document}
\begin{center}
\centerline{\large\bf Diffusion in scale-free networks with annealed disorder}

\bigskip

Dietrich Stauffer$^\dagger$ and Muhammad Sahimi

{\it Department of Chemical Engineering, University of Southern California,
Los Angeles, California 90089-1211, USA}

\end{center}

The scale-free (SF) networks that have been studied so far contained {\it 
quenched} disorder generated by random dilution which does not vary with the
time. In practice, if a SF network is to represent, for example, the 
worldwide web, then the links between its various nodes may temporarily be 
lost, and re-established again later on. This gives rise to SF networks with 
{\it annealed} disorder. Even if the disorder is quenched, it may be more
realistic to generate it by a dynamical process that is happening in the 
network. In this paper, we study diffusion in SF networks with annealed 
disorder generated by various scenarios, as well as in SF networks with 
quenched disorder which, however, is generated by the diffusion process itself.
Several quantities of the diffusion process are computed, including the mean
number of distinct sites visited, the mean number of returns to the origin, and
the mean number of connected nodes that are accessible to the random walkers at
any given time. The results including, (1) greatly reduced growth with the time
of the mean number of distinct sites visited; (2) blocking of the random
walkers; (3) the existence of a phase diagram that separates the region in
which diffusion is possible from one in which diffusion is impossible, and (4)
a transition in the structure of the networks at which the mean number of
distinct sites visited vanishes, indicate completely different behavior for the
computed quantities than those in SF networks with quenched disorder generated
by simple random dilution.

\bigskip

\noindent PACS number(s): 89.75.Da, 05.40.Fb, 82.20.-w

\bigskip

\noindent\rule{3truecm}{0.01in}

\noindent $^\dagger$ Permanent address: Institute for Theoretical Physics, 
Cologne University, D-50923 K\"oln, Germany.

\newpage

\begin{center}
{\bf I. INTRODUCTION}
\end{center}

Scale-free (SF) networks have recently attracted much interest [1,2]. In such
networks the probability distribution $f(k)$ for a node to have $k$ links to 
other nodes follows a power law,
\begin{equation}
f(k) \sim k^{-\gamma}\;,
\end{equation}
where $\gamma$ is a parameter that measures how well-connected the networks 
are. Many unusual properties of SF networks have been attributed to the 
distribution (1). For example, if $2\leq\gamma\leq 3$, then such networks 
preserve their connected structure even if their nodes are randomly removed and
their structure is greatly diluted, indicating their resilience to random 
failure, while they exhibit vulnerability to correlated dilution whereby only 
the most connected nodes are removed [3-7]. In addition, Gallos and Argyrakis 
[8] showed that certain chemical reactions in SF networks exhibit behavior 
drastically different from the same reactions in regular lattices. Many 
properties of SF networks have been computed, including their metric [9] and 
percolation properties [10]. Moreover, such phenomena as epidemic spreading 
[11], with the intended applications being the spread of computer virus and 
pollution control, and the spread of extreme opinions advocated by fanatics in 
a society have also been studied [12] in SF networks. Most of the previous 
studies of SF networks employed computer simulations in order to compute 
various properties of interest. Hughes, Reed, and co-workers [13] developed a 
class of networks whose properties can be computed either exactly or by 
accurate mean-field approximations.

Transport processes occurring on SF networks are also of interest. For example,
if SF networks are supposed to model the internet, then diffusion of particles
on such networks may represent a person trying to locate information by 
visiting a website and its direct and indirect links, or the random spreading 
of a virus throughout the internet. Alternatively, such a process might also
be relevant to the spread of a disease in biological networks. Diffusion (and
reaction) processes have already been studied in many different systems 
[14-16], but studies of the properties of such processes in SF networks have
started only recently [17-20]. A recent summary is given by Bollt and 
ben-Avraham [21].

All the studies mentioned above considered SF networks with {\it quenched}
disorder. That is, once the disorder is generated in the networks, either 
through the power-law distribution of the connectivities, Eq. (1), or by 
random dilution, it remains fixed. In practice, {\it annealed} disorder - one 
that changes with the time - may be more relevant. As an example, consider the 
worldwide web (WWW). At any given time, the links between a number of sites may
temporarily be lost due to, for example, some of the sites being blocked, shut 
down, or hacked. The links may also be lost over a period of time because, for
example, certain sites are visited so often that it becomes very difficult for 
a new person to visit them. Such disconnections can, at any given time, happen 
between a number of sites, hence giving rise to annealed disorder whereby the 
existence of links between various nodes may change with the time. 

Even if the disorder is quenched, for application to the WWW it may be more 
reasonable if it is generated by a dynamical process which is happening in the 
network. For example, if a virus is spreading throughout the WWW, the sites 
that are visited by the virus and have become disabled must be removed (either 
temporarily or permanently) from the network. Therefore, if, for example, 
diffusion of particles on SF networks can represent the spread of a virus in 
the WWW, then studying diffusion in SF networks with annealed disorder, or 
quenched disorder which is generated by the random walk itself, may be more 
relevant to the problem. 

The purpose of the present paper is to carry out a study of random walk 
processes in SF networks in which the disorder is either annealed, or if it is 
quenched it is generated by the random walk process itself. We present the 
results of extensive Monte Carlo simulations of diffusion in such SF networks 
which, to our knowledge, has never been studied before. We consider several 
scenarios for generating the disorder, and compute a number of important 
properties of diffusion in SF networks. Diffusion in regular lattices with 
annealed disorder was studied a long time ago [22].

The plan of this paper is as follows. In the next section we describe models
of annealed disorder that we consider. Section III describes the Monte Carlo
simulation procedure and the quantities that we compute. The results are
presented and discussed in Section IV.

\begin{center}
{\bf II. MODELS OF DISORDER}
\end{center}

Before introducing any disorder into a SF network, we grow the network
with $N$ sites such that every new site selects four of the already existing 
sites as neighbors, with a probability proportional to the number of neighbors
the selected site had before. We start with four sites connected to each other.
Such a method for generating a SF network has been used in numerous papers [1].
In such a network, the probability for a site to have $k\ge 4$ neighbours 
decays as $1/k^3$ (i.e., $\gamma=3$). We do not attempt to vary $\gamma$ as a 
parameter of the model. We now consider several scenarios under which the 
topology of the grown network may change with the time. These are as follows.

\begin{center}
{\bf Model A}
\end{center}

In the first model we let the SF network change under the influence of the 
walkers, say a virus that is penetrating the WWW. At each time step, every 
site which is not yet visited is deleted permanently from the network with 
probability $p$. The idea is that if, for example, a virus is spreading 
throughout the WWW, at every time step, a certain fraction of the uninfected 
sites becomes aware of the existence of the virus, and take themselves out of 
the WWW for a certain time. Alternatively, we also study the case in which the 
already visited sites (representing the infected sites) are deleted in order to
compare the results with the first scenario.

\begin{center}
{\bf Model B}
\end{center}

In this model only after visiting a site (of the previously constructed 
network) is attempted $m$ times, the site becomes available permanently to the
walkers. The idea is that, as it often happens in practice, a site cannot
be accessed by a visitor because, for example, it has been shut down for a
certain period of time, or it is already being visited so heavily that it
becomes much more difficult for additional visitors to access the site, or that
one has forgotten the precise address of a site and, therefore, many attempts 
are made in order to find the correct address of the site to access it.

\begin{center}
{\bf Model C}
\end{center}

In this model we combine the two effects of forgetting a site and having to 
try repeatedly, in order to generate annealed disorder.

\begin{center}
{\bf Model D}
\end{center}

In this model a random walker still needs, as in model C, $m$ attempts to visit
a new site, but if a site is visited, then a repeat visit is possible only if 
the total number of attempts is at least $m+P\times\min(m,\Delta t)$, where 
$\Delta t$ (measured in units of the number of steps taken) is the time elapsed
after the last successful visit, and $P\ll 1$ a new "forgetfulness" parameter.
Thus, a site which is visited, but then not visited again for some time 
(which can be long), is treated as if it was never visited before which, from a
practical view point regarding the WWW, seems reasonable. Hence, this type of 
annealed disorder involves a feedback mechanism between consecutive visits to 
the sites of a SF network, or the WWW.

\begin{center}
{\bf Model E}
\end{center}

This model is motivated by the fact that, if the best connected sites of a SF
network are removed, most of the network's connectivity is lost [3-7]. Thus,
we first remove a certain fraction of the better connected sites of the 
network, and then use Model D in order to generate annealed disorder and carry 
out simulation of the diffusion process. Therefore, Model E contains both
quenched and annealed disorder.

\begin{center} 
{\bf III. MONTE CARLO SIMULATION}
\end{center}

A property of random walk processes, which from a practical point of view 
is most relevant to SF networks and their possible applications to WWW, is the 
mean number of distinct sites visited after time $t$, $S(t)$, which 
characterizes the coverage of the space by the random walkers. The SF networks 
that we utilize are not embedded in Euclidean space. In addition, it has been 
shown (see Cohen and Havlin [9]) that the diameter of SF networks that we
simulate (with $\gamma=3$), which represents the minimal number of links needed
to connect two network sites, is only of the order of $\ln N$, so that, for 
example, the WWW with about $8\times 10^8$ nodes has a very small diameter 
[23]. Therefore, the mean-square displacement of the random walkers is not a 
very useful property, because it does not measure any true distance that the 
random walkers travel in the space.

For some of the models we also compute the number $R(t)$ of returns to the 
origin of the random walks. The traditional probability $P_0(t)$ that the 
random walkers return to the origin of their walks is simply $\propto 
dR(t)/dt$. Also 
computed is the number $A(t)$ of the sites that are accessible to the random
walkers at time $t$. The latter quantity is particularly important to the 
application involving WWW and the spread of a virus in computer networks. Most 
of our results were obtained using $10^2$ or $10^3$ realizations of the SF 
networks with up to $N=10^6$ sites each. 

\begin{center}
{\bf IV. RESULTS AND DISCUSSIONS}
\end{center}

In the first series of simulations, we put one random walker onto one of the 
four initial sites of the SF network, and at each time step $t$ the walker 
selects with probability 1/2 one of its $k$ neighbours and moves there. Each 
connection between two neighboring sites can be travelled in both directions. 
If, instead, we take directed or hierarchical SF networks with one-way links 
(so that a hop from one site to another is allowed only in a fixed direction), 
the walkers would soon be trapped. We average for each realization the results 
over the four starting points, and then over all network realizations. 

Figure 1 (top) indicates that the mean number of distinct sites visited is 
given by
\begin{equation}
S(t)\propto t\;,
\end{equation}
for intermediate times (see also Gallos [18]). Equation (2) is not valid for 
short times. For long times finite-size effects dominate and one obtains $S=N$,
i.e., every site of the network will eventually be visited. The fraction 
$S(t)/N$ of the visited sites is a size-independent function of $t/N$ for not 
too short times. This is shown in Fig. 1 (bottom) which indicates that 
\begin{equation}
S(t)/N=f(t/N)\;.
\end{equation}
Scaling law (3) is analogous to opinion dynamics [24], where the number $S(t)$
of surviving opinions equals the number $N$ of possible opinions, multiplied
by a scaling function $f(t/N)$, where $t$ represents the number of people. 
The linear dependence of $S(t)$ on $t$ is similar to what one obtains for
random walks on the Bethe lattices or Cayley trees, branching structures in 
which every node is connected to a fixed number $k$ of other nodes with no 
closed loops allowed. It was shown by Hughes and Sahimi that for such trees 
[25]
\begin{equation}
S(t) \sim \frac{k-2}{k-1}\;t\;.
\end{equation}
Hughes, Sahimi, and Davis [25] showed that even if small closed loops are 
allowed to form in the tree structure, the essence of Eq. (4), namely, the 
linear dependence of $S(t)$ on $t$, would remain unaltered.

Figure 2 shows the results for $S(t)$, computed for Model A, for several
values of the blocking probability $p$. For better efficiency in the 
simulations we delete only sites which are just visited by the ant, with 
probability $pt$ instead of $1-(1-p)^t$. Thus, at time $t=1/p$ the annealed 
disorder no longer has any effect and the network is quenched. The results 
indicate that the annealed disorder generated by blocking has an effect similar
to {\it a priori} restriction of the network {\it size} shown in Fig. 1. 
It does not matter much whether all sites, or only previously unvisited sites,
can be removed.

Figure 3 presents the results for $S(t)$, computed for Model B, using several
values of the parameters $m$, the number of attempts to visit a site before
that site becomes available. The results indicate the strong delay in the 
number of distinct sites visited, especially for large values of $m$. In 
particular, for short and intermediate times $S(t)$ varies with $t$ nonlinearly
and in a nontrivial manner. Eventually, however, the linear growth of $S(t)$ 
with $t$ seems to set in. 

The results for $S(t)$ for Model C, in which we combine the disorder in Models
A and B, are shown in Fig. 4 (top). Even for an extremely small and fixed 
value, $p=10^{-5}$, diffusion becomes increasingly difficult as $m$ increases. 
Strong delays in the growth of $S(t)$ are seen even for $m$ as small as 5. For 
$m=50$, the largest value that we considered, there is a long delay in the 
growth of $S(t)$ with $t$, followed by a nonlinear growth. At long times, 
$S(t)$ more or less saturates at values much smaller than $N$, the total number
of sites in the network. Similarly, for a fixed $m$ as small as 2, diffusion 
becomes impossible for $p$ as small $0.1$. Even for $p=10^{-2}$ the walker
visits very few sites before its motion becomes incapable of taking it to new
sites. These results are also shown in Fig. 4 (bottom).

Figure 5 presents a phase diagram for Model D, indicating that for a certain 
region in the ($m,P$) space, diffusion becomes impossible and the walkers 
become blocked. The curve that separates the diffusion/no diffusion regions is 
given roughly by, $m=1/P$, so that for $m>1/P$ no diffusion is possible and the
random walkers remain at their starting place. For $m<1/P$ the results agree 
with those for $P=0$, when the model becomes equivalent to Model B.

Figure 6, showing the average number $R(t)$ 
of visits to the origin, indicates that in
Model D the random walkers return to their starting points very rarely. 
Even at relatively long times, very few visits to the origin take place and,
in fact, with increasing $m$ the visits become more rare, since more sites
are unavailable at any given time. The traditional probability $P_0(t)$ of
return to the origin that has been calculated for many random walks is simply 
the derivative with respect to $t$ of the curves shown in Fig. 6, which would 
indicate a sharp decay of $P_0(t)$ with $t$ (not shown).

An important quantity is the number of nodes that are accessible to a random 
walker during its motion. For example, if a virus is moving through the WWW,
the number of nodes that it can potentially reach in order to infect them 
determines the "success" of the virus. We have computed this quantity for Model
D; the results are shown in Fig. 7 (top), where we show the dependence of the 
number of accessible sites on the time $t$ for several network sizes, with
$P=0.1$ and $m=5$. Qualitatively, the results are similar to the number of 
distinct sites visited, although numerically they are quite different. For 
intermediate times, the number of accessible sites follows a scaling law 
similar to Fig. 1 for the number of visited sites. This is demonstrated in 
Figure 7 (bottom).

Finally, for Model E, we remove permanently a fraction $q$ of the $N$ network 
sites and study, using Model D, how many sites can still be visited for $t\to
\infty$. If the deletion is random, $q$ must be close to unity to split the 
network into small parts [3]; if the $qN$ most connected sites are removed, 
then a small $q$ is sufficient to split the network [5]. We consider none of 
such extreme limits, but instead remove the first $qN$ sites which joined the 
network during its creation. They are usually the most connected ones, but 
exceptions do exist. Moreover, the initial core of four site belongs to the 
$qN$ deleted sites. Thus, the random walkers start their diffusion on such 
sites but can never return there. Figure 8 suggests a phase transition at 
intermediate values of $q$, from a large connected remaining network at small 
$q$, to a fragmented assembly of much smaller clusters at large $q$, as in 
percolation on square lattices. Similar results, but with much less precise 
statistics, were obtained for the number of accessible sites (not shown).

Let us emphasize that none of the results presented in this paper are obtained 
when one considers random walks in SF networks with quenched disorder generated
by simple random dilution (as opposed to being generated by the dynamical
process itself). We believe that the results presented in the present paper 
indicate not only the significance of the disorder of the type we consider, 
but also their potential applicability to networks encountered in practice, and
in particular the WWW which contains either annealed disorder, or quenched 
disorder which is, however, generated by a dynamical process.

We thank L. Gallos for helpful information.

\begin{center}
{\bf V. SUMMARY}
\end{center}

We carried out extensive Monte Carlo simulation of diffusion in scale-free
networks with either annealed disorder, or quenched disorder which, however,
is generated by the diffusion process itself. Several models of disorder, 
motivated by application to the worldwide web, were considered and important 
properties of the diffusion process, such as the number of distinct sites 
visited and the fraction of accessible sites to the walkers at any given time, 
were computed. The results indicate their drastic departure from those in SF 
networks with quenched disorder generated by simple random dilution. For some 
of the models we considered, scaling laws for the effects of finite times and 
finite sizes were demonstrated. For the more complex models of disorder, where 
the structure of the available network is determined by the diffusion process 
itself, diffusion may become impossible in some parameter range, hence 
providing potential strategies to handle certain phenomena in SF networks and 
the WWW, such the spread of a virus in such networks.

\newpage

\newcounter{bean}
\begin{list}%
{[\arabic{bean}]}{\usecounter{bean}\setlength{\rightmargin}{\leftmargin}}

\item R. Albert and A.L. Barab\'asi, Rev. Mod. Phys. {\bf 74}, 47 (2002).

\item J.F.F. Mendes and S.N. Dorogovtsev, {\it Evolution of Networks: From
Biological Nets to the Internet and the WWW} (Oxford University Press, London,
2003).

\item R. Cohen, K. Erez, D. ben-Avraham, and S. Havlin, Phys. Rev. Lett.
{\bf 85}, 4626 (2000).

\item R. Albert, H. Jeong, and A.-L. Barab\'asi, Nature (London) {\bf 406},
378 (2000).

\item R. Cohen, K. Erez, D. ben-Avraham, and S. Havlin, Phys. Rev. Lett. {\bf 
86}, 3682 (2001).

\item Y. Moreno, J.B. G\'omez, and A.F. Pacheco, Europhys. Lett. {\bf 58},
630 (2002).

\item L. Gallos, R. Cohen, P. Argyrakis, A. Bunde, and S. Havlin, Phys. Rev.
Lett. {\bf 94}, 188701 (2005). 

\item L. Gallos and P. Argyrakis, Phys. Rev. Lett. {\bf 92}, 138301 (2004).

\item A.F. Rozenfeld, R. Cohen, D. ben-Avraham, and S. Havlin, Phys. Rev. Lett.
{\bf 89}, 218701 (2002); C.P. Warren, L.M. Sander, and I.M. Sokolov, Phys. Rev.
E {\bf 66}, 056105 (2002); R. Cohen and S. Havlin, Phys. Rev. Lett. {\bf 90},
058701 (2003); G. Bianconi and A. Capocci, Phys. Rev. Lett. {\bf 90}, 078701
(2003).

\item S.N. Dorogovtsev, J.F.F. Mendes, and A.N. Samukhin, Phys. Rev. E {\bf 
64}, 066110 (2001); N. Schwartz, R. Cohen, D. ben-Avraham, A.-L. Barab\'asi,
and S. Havlin, {\it ibid.} {\bf 66}, 015104 (2002); R. Cohen, D. ben-Avraham, 
and S. Havlin, {\it ibid.} {\bf 66}, 036113 (2002).

\item R. Pastor-Satorras and A. Vespignani, Phys. Rev. Lett. {\bf 86}, 3200
(2001); D. Volchenkov, L. Volchenkova, and Ph. Blanchard, Phys. Rev. E {\bf
66}, 046137 (2002); L.K. Gallos and P. Argyrakis, Physica A {\bf 330}, 117
(2003).

\item D. Stauffer and M. Sahimi (unpublished), physics/0506106 and 0506154.

\item W.J. Reed and B.D. Hughes, J. Theor. Biol. {\bf 217}, 125 (2002); Phys. 
Rev. E {\bf 66}, 067103 (2002); Physica A {\bf 319}, 579 (2003); D.Y.C. Chan, 
B.D. Hughes, A.S. Leong, and W.J. Reed, Phys. Rev. E {\bf 68}, 066124 (2003).

\item B.D. Hughes, {\it Random Walks and Random Environments}, Volume 1
(Oxford University Press, London, 1995).

\item D. ben-Avraham and S. Havlin, {\it Diffusion and Reactions in Fractals
and Disordered Systems} (Combridge University Press, Cambridge, 2000).

\item M. Sahimi, {\it Heterogeneous Materials I} (Springer, New York, 2003), 
Chapters 5 and 6.

\item J.D. Noh and H. Rieger, Phys. Rev. Lett. {\bf 92}, 118701 (2004).

\item L.K. Gallos, Phys. Rev. E {\bf 70}, 046116 (2004).

\item T. Di Matteo, T. Aste, and M. Gallegati, physics/0406091.

\item I. Simonsen, K.A. Eriksen, S. Maslov, and K. Sneppen, Physica A 
{\bf 336}, 163 (2004).

\item E.M. Bollt and D. ben-Avraham, New J. Phys. {\bf 7}, paper 26 (2005). 

\item M. Sahimi, J. Phys. C {\bf 19}, 1311 (1986).

\item R. Albert, H. Jeong, and A.-L. Barab\'asi, Nature (London) {\bf 401},
130 (1999).

\item D. Stauffer, A.O. Sousa, and C. Schulze, J. Artificial Societies and 
Social Simulation (jasss.soc.surrey.ac.uk) {\bf 7} (No. 3), paper 7; S. 
Fortunato, Int. J. Mod. Phys. C {\bf 15}, 1021 (2004); F. A. Rodrigues and L. 
da F. Costa, Int. J. Mod. Phys. C {\bf 16} (No. 11) (2005).

\item B.D. Hughes and M. Sahimi, J. Stat. Phys. {\bf 29}, 781 (1982); B.D.
Hughes, M. Sahimi, and H.T. Davis, Physica A {\bf 120}, 515 (1983).

\end{list}%

\begin{figure}[hbt]
\begin{center}
\includegraphics[angle=-90,scale=0.35]{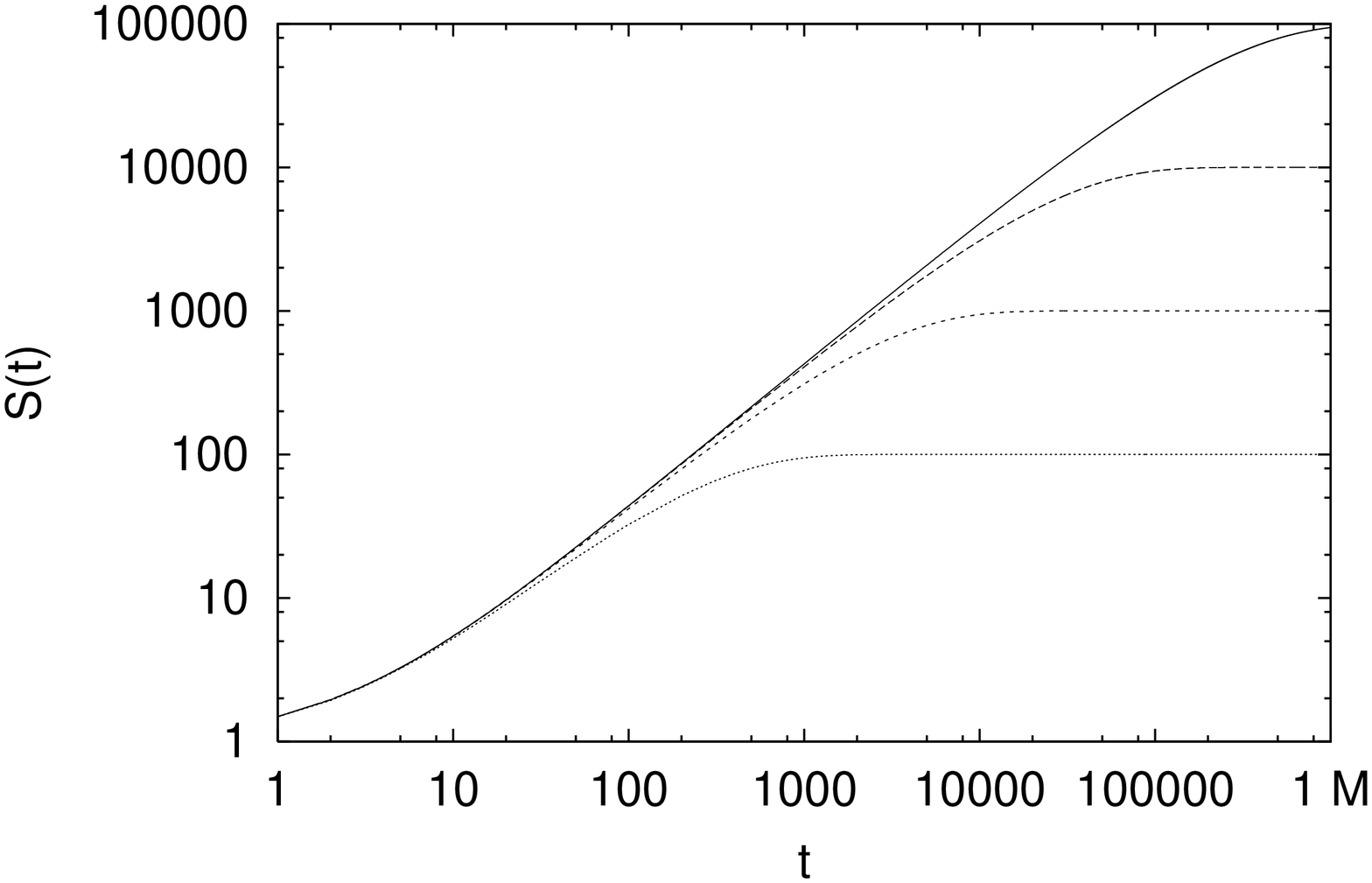}
\includegraphics[angle=-90,scale=0.35]{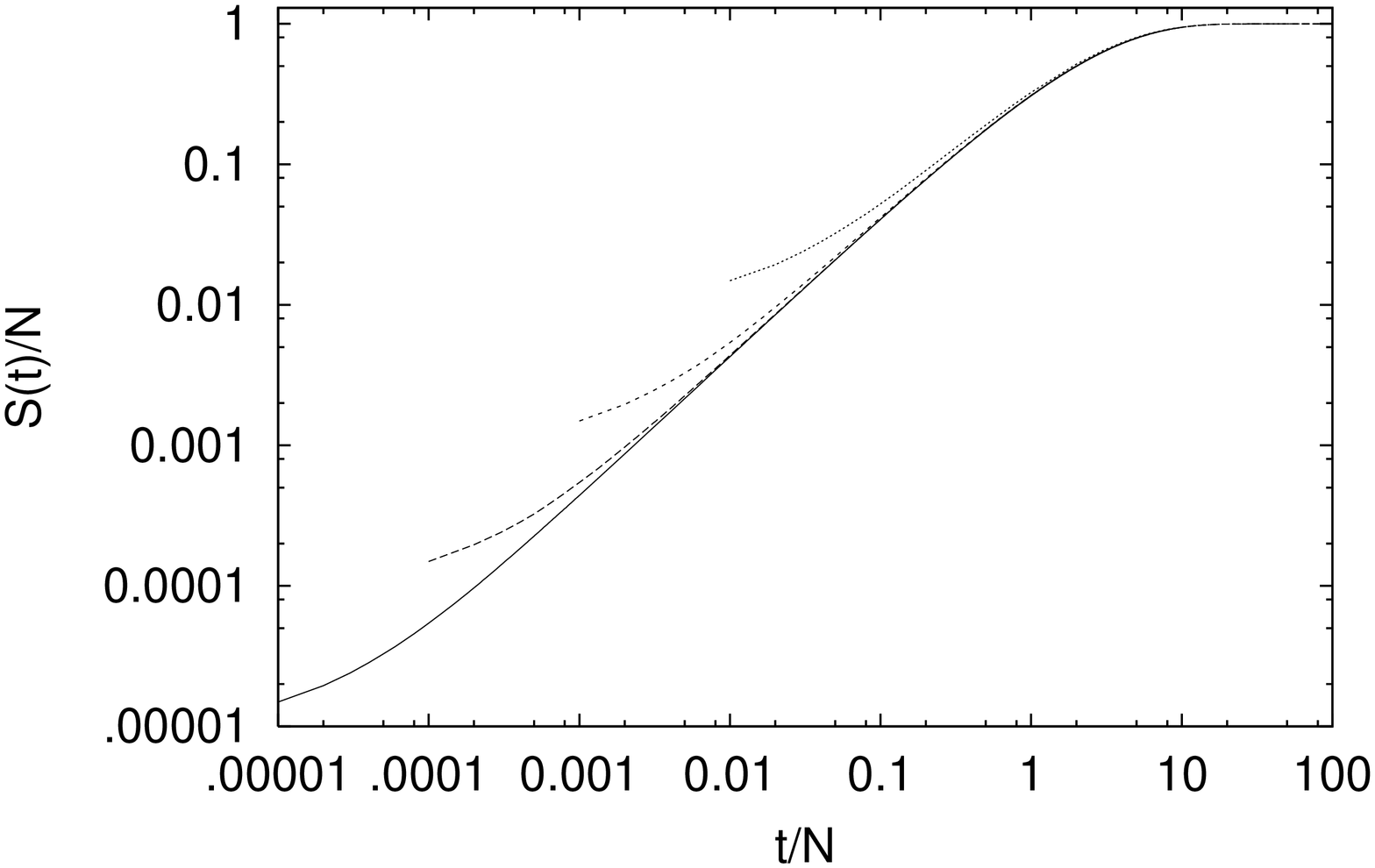}
\end{center}
\caption{Top: Time-dependence of the number of distinct sites visited, averaged 
over $10^3$ samples of four walkers each. The static network has, from top to
bottom, $N=10^5,\;10^4,\;10^3,$ and $10^2$ sites. Bottom: Scaling 
representation of the results in the top figure, with $N$ increasing from right
to left.}
\end{figure}

\begin{figure}[hbt]
\begin{center}
\includegraphics[angle=-90,scale=0.35]{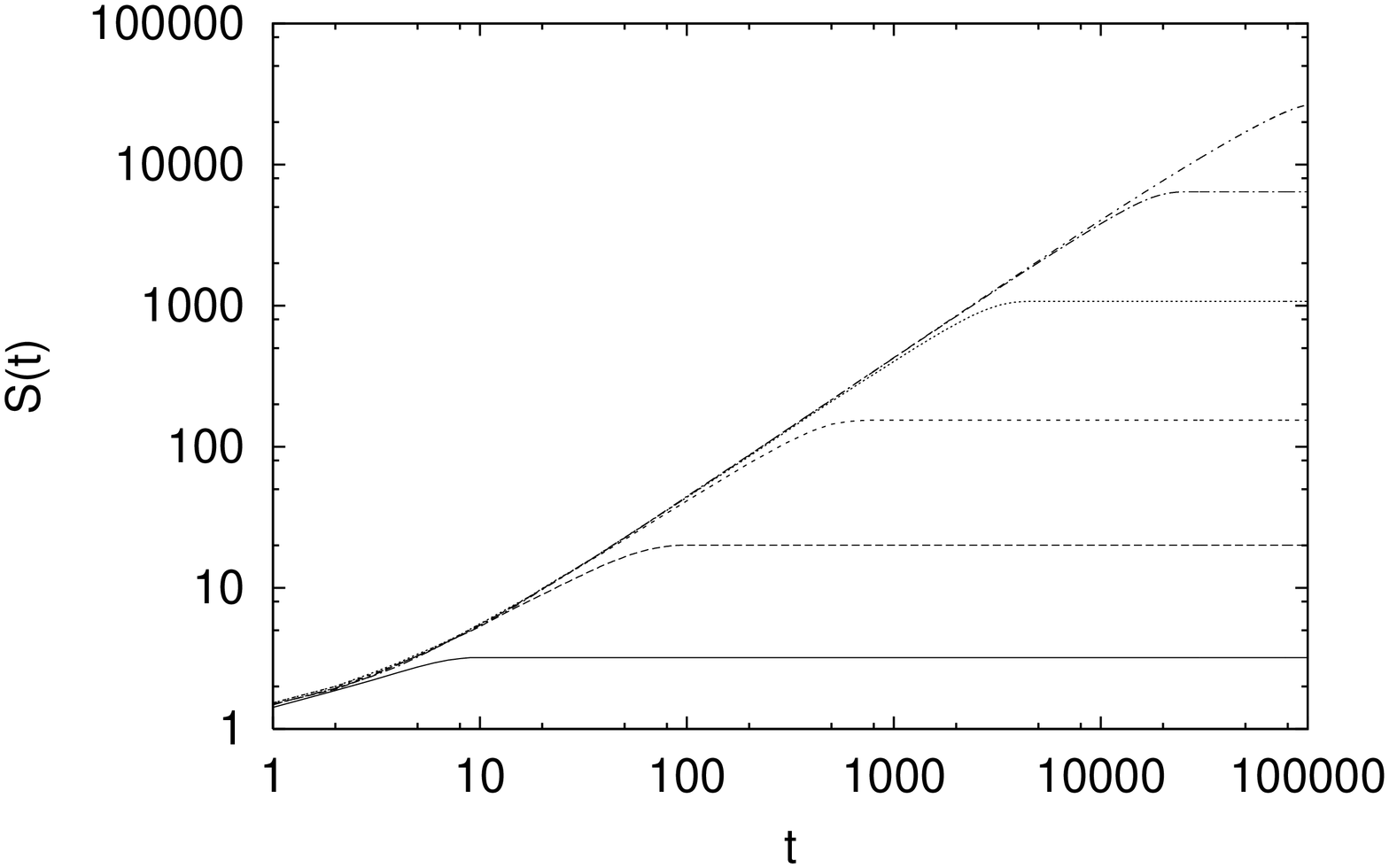}
\includegraphics[angle=-90,scale=0.35]{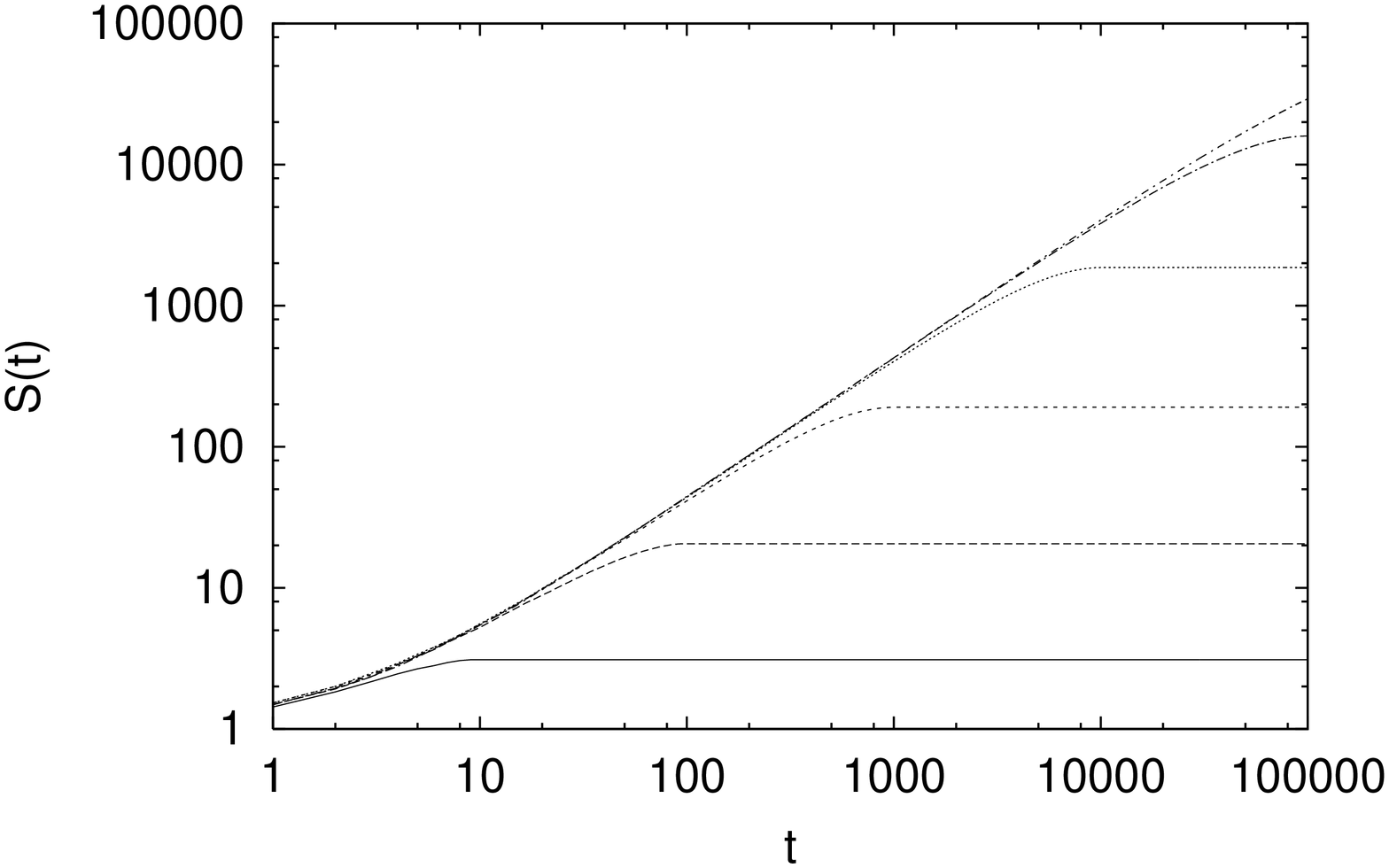}
\end{center}
\caption{Number of distinct sites visited in Model A as a function of time $t$,
averaged over 100 samples of four walkers each. The network contains 
disorder with a fraction $pt$ of the sites not available at time $t$ with,
from top to bottom, $p=10^{-6},\;10^{-5},\;10^{-4},\;10^{-3},\;10^{-2}$, and
$10^{-1}$.} In the top part, all sites can be deleted, in the bottom part only
the unvisited sites.
\end{figure}

\begin{figure}[hbt]
\begin{center}
\includegraphics[angle=-90,scale=0.5]{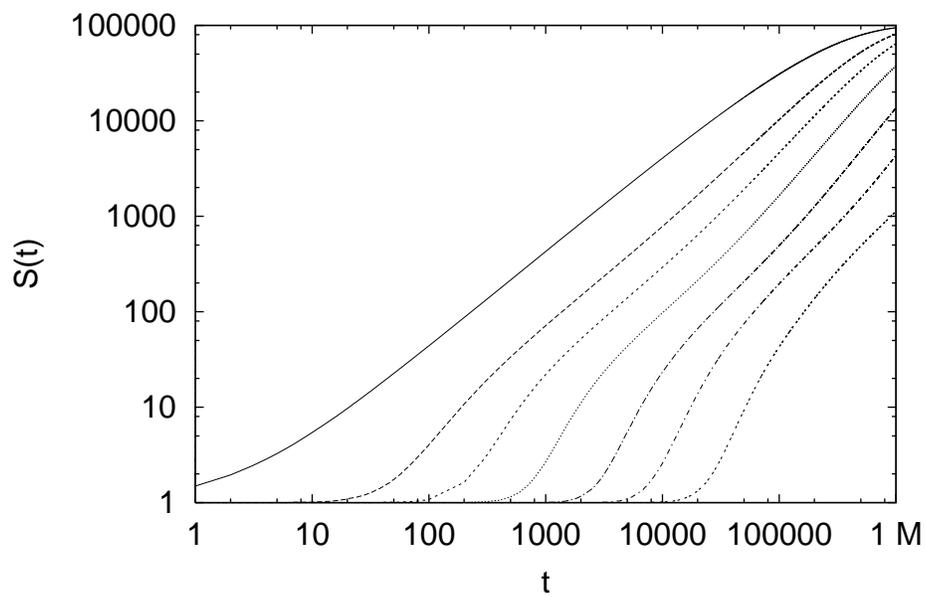}
\end{center}
\caption{Number of distinct sites visited in Model B. The results are for, from
left to right, $m=1$ (same as in Fig. 1), 2, 3, 5, 10, 20, and 50, all computed
for networks of size $N=10^5$ and averaged over $10^3$ samples.}
\end{figure}

\begin{figure}[hbt]
\begin{center}
\includegraphics[angle=-90,scale=0.35]{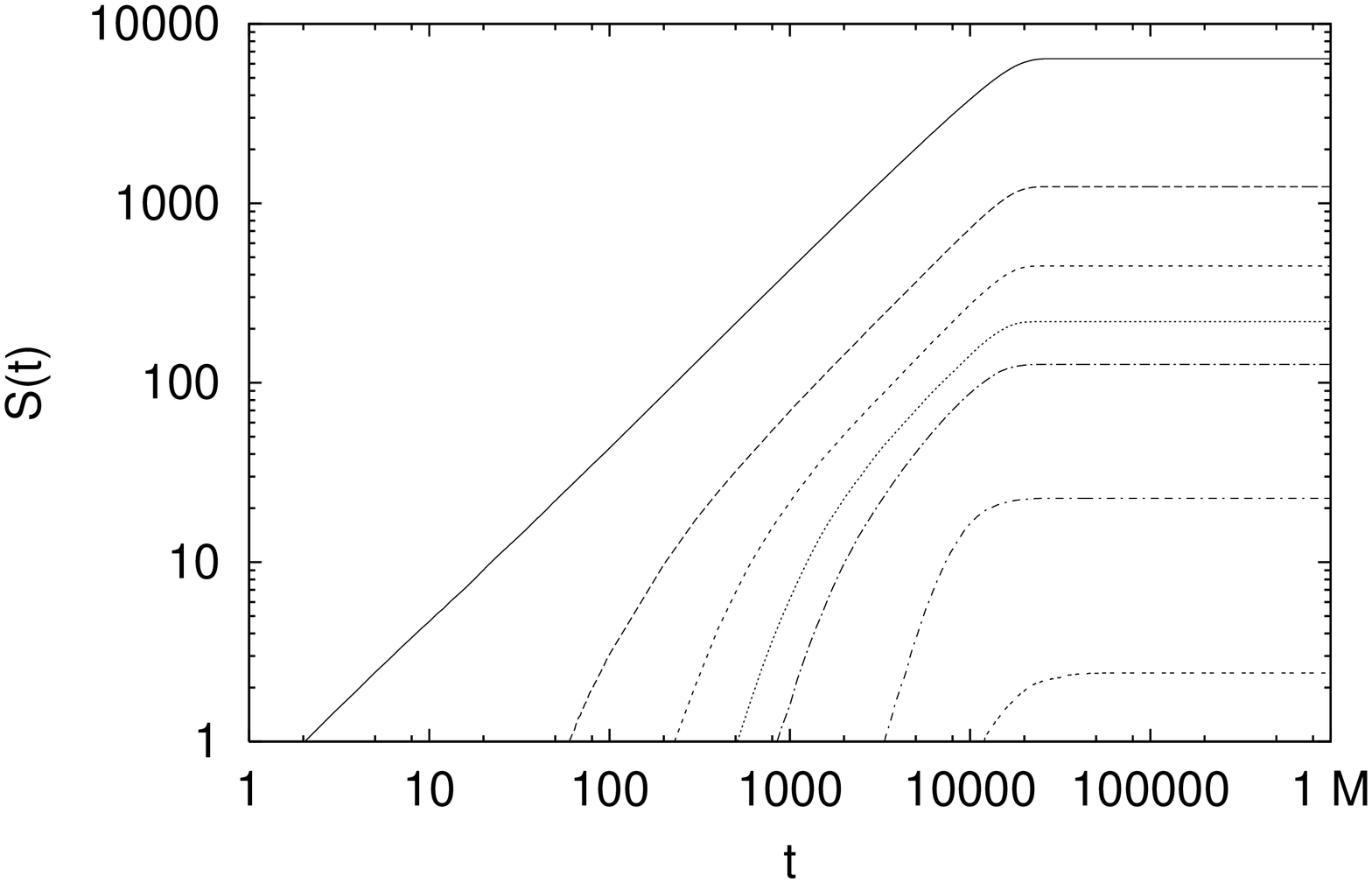}
\includegraphics[angle=-90,scale=0.35]{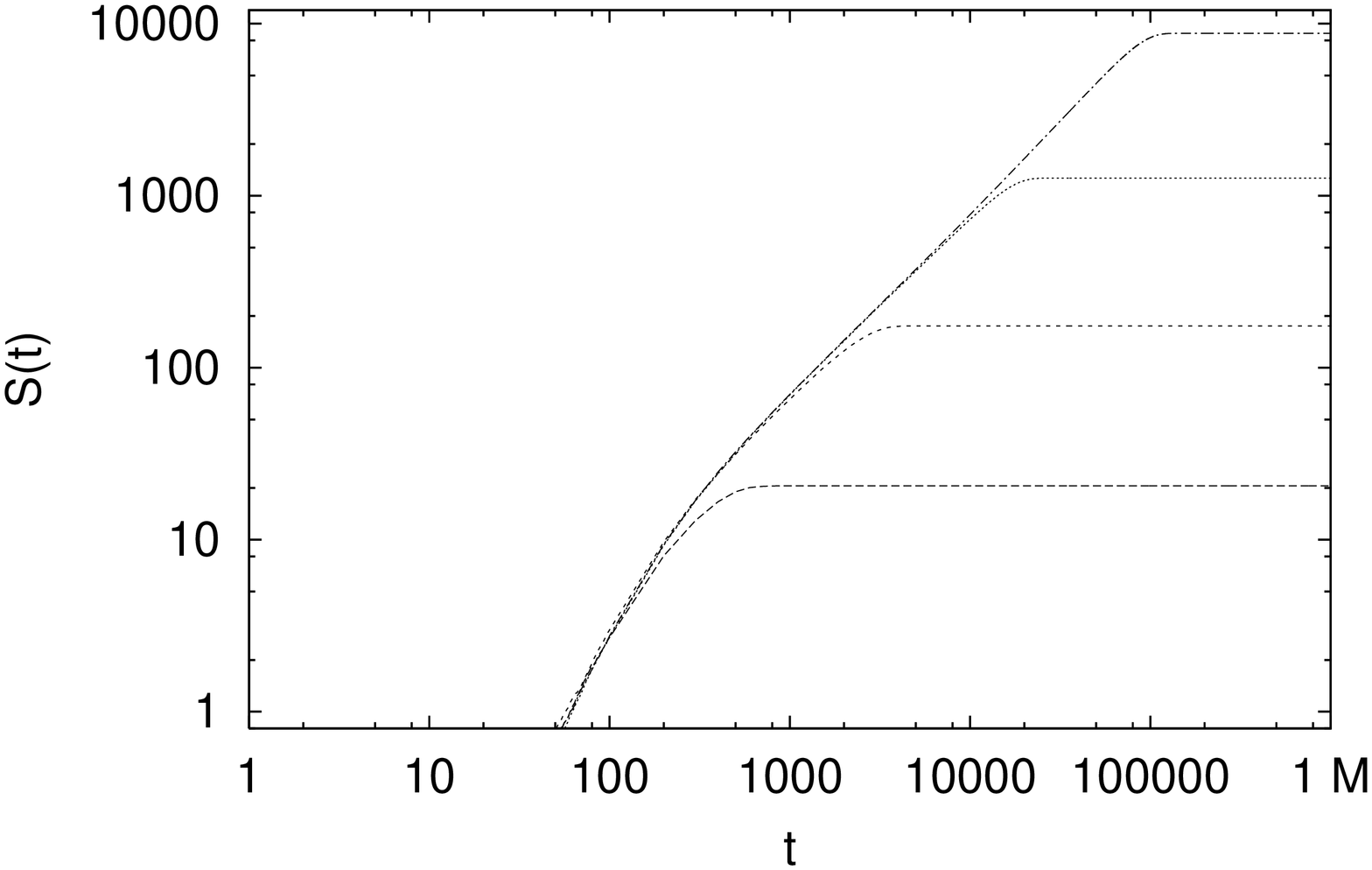}
\end{center}
\caption{Top: Number of distinct sites visited, averaged over 100 samples, in
Model C with $p=10^{-5}$ and, from left to right, $m= 1$, 2, 3, 4, 5, 10, 
and 20. (The results for $m=50$ are too small to be seen; here we do
not count the starting point as being visited at time zero.)
Bottom: Results for, from bottom to top, $p=10^{-3},\;
10^{-4},\; 10^{-5},$ and $10^{-6}$ and $m=2$. For $p\ge 0.1$ no diffusion was 
possible, while $S(t) < 1$ for $p=0.01$.}
\end{figure}

\begin{figure}[hbt]
\begin{center}
\includegraphics[angle=-90,scale=0.5]{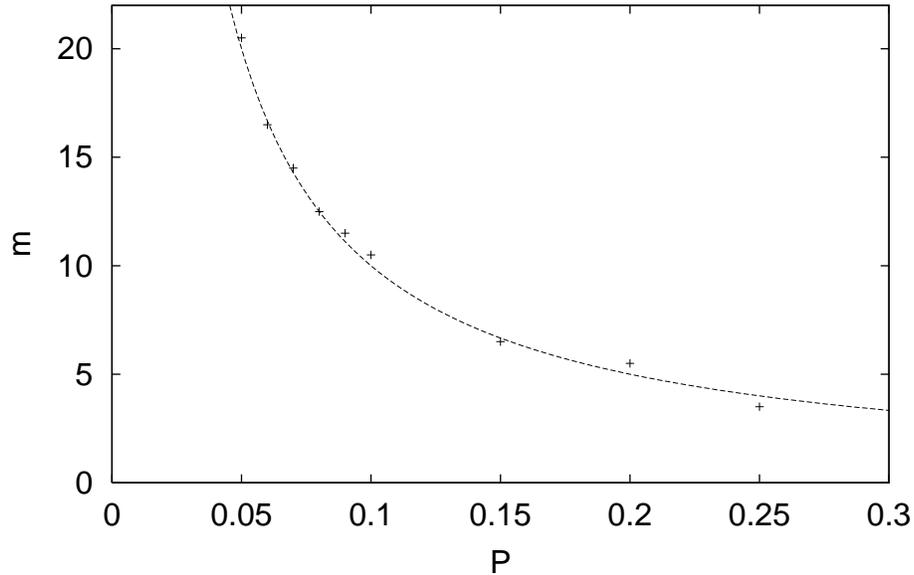}
\end{center}
\caption{Phase diagram for Model D. For pairs ($m,P$) that are above the curve
diffusion is impossible. The results are averages over 10 or 100 realizations 
of $N = 10^4$. The curve shown is $m=1/P$.}
\end{figure}

\begin{figure}[hbt]
\begin{center}
\includegraphics[angle=-90,scale=0.5]{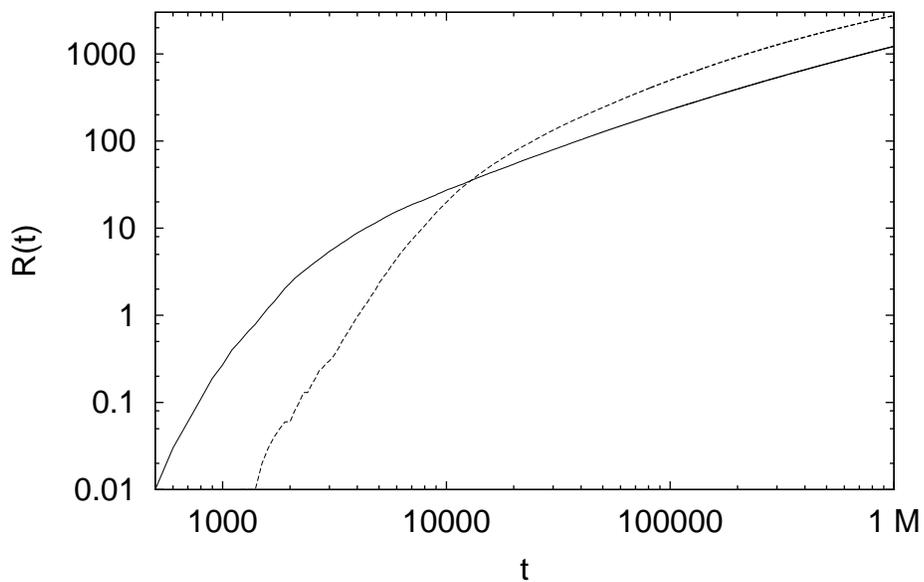}
\end{center}
\caption{Total number of visits to the origin, up to the time $t$, in Model D.
The results are averages over 100 realizations for $P=0.1$ and $m=5$ (solid
line) and 10 (dashed line).}
\end{figure}

\begin{figure}[hbt]
\begin{center}
\includegraphics[angle=-90,scale=0.35]{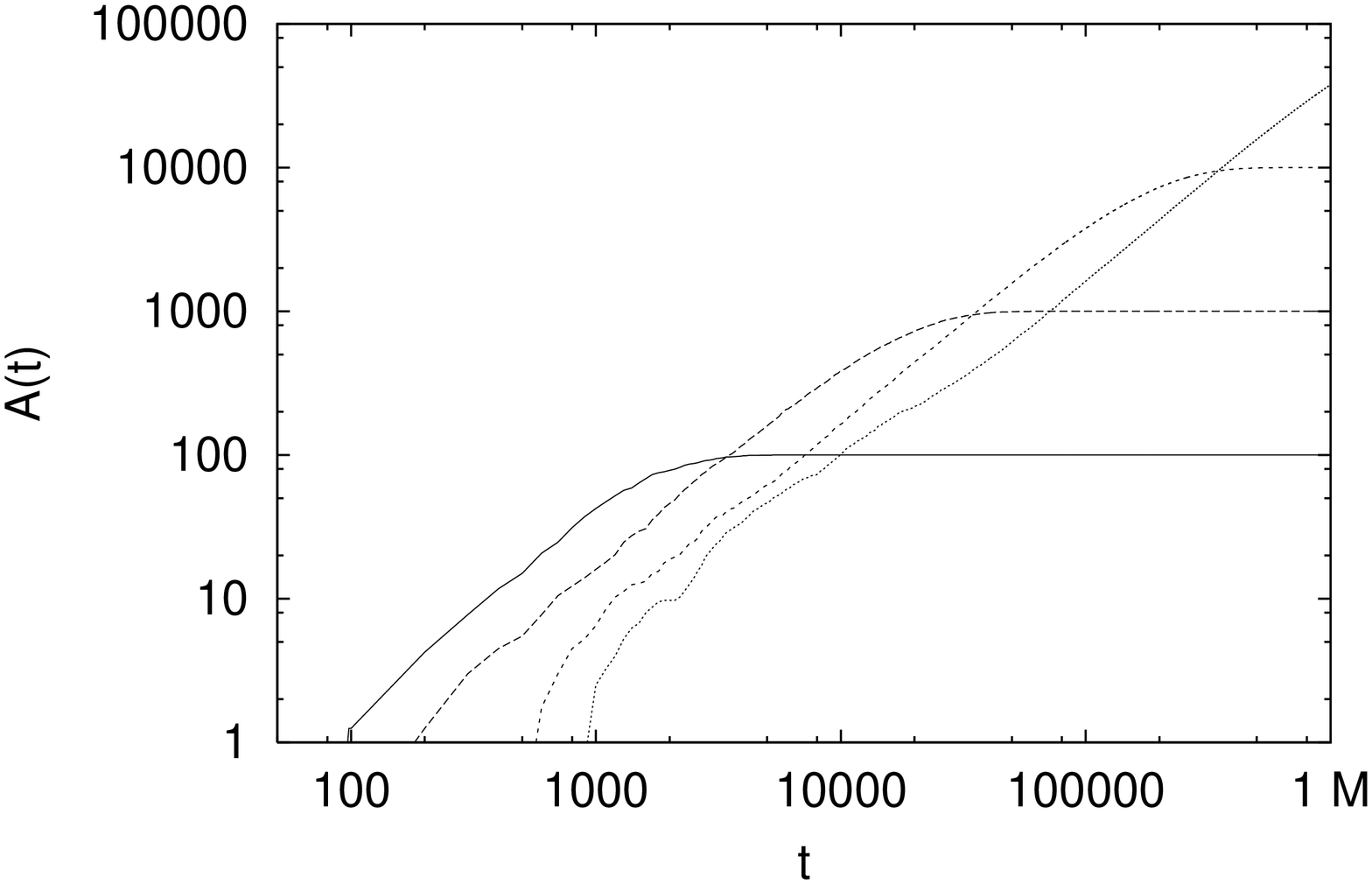}
\includegraphics[angle=-90,scale=0.35]{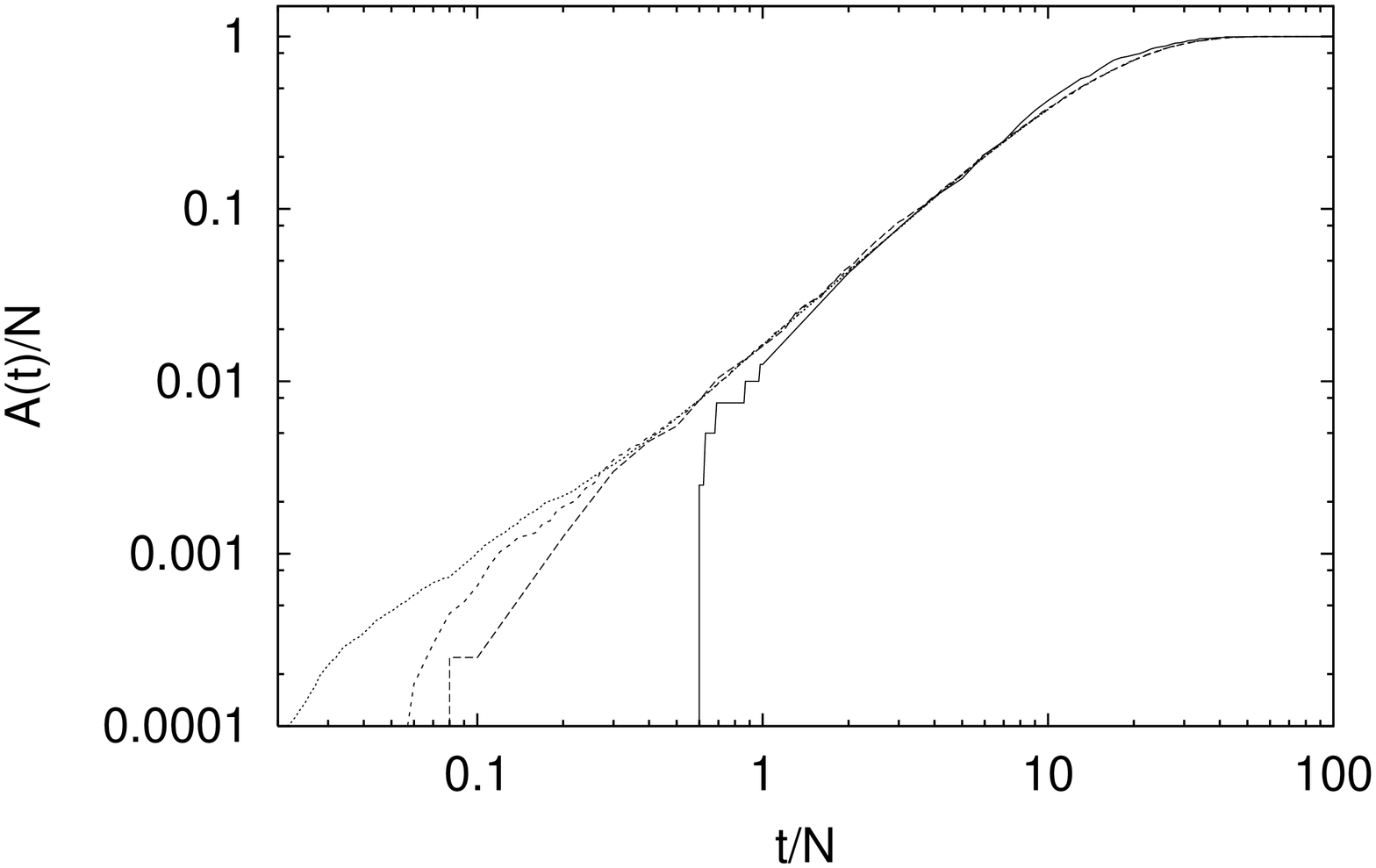}
\end{center}
\caption{Top: Number of sites that are accessible to the random walkers at 
time $t$ in Model D for, from bottom to top, network sizes $N=10^2,\;10^3,\;
10^4,$ and $10^5$ (one realization with four walkers). Bottom: Scaled results of
those shown at the top. Both figures are for $P=0.1$ and $m=5$.}
\end{figure}

\begin{figure}[hbt]
\begin{center}
\includegraphics[angle=-90,scale=0.5]{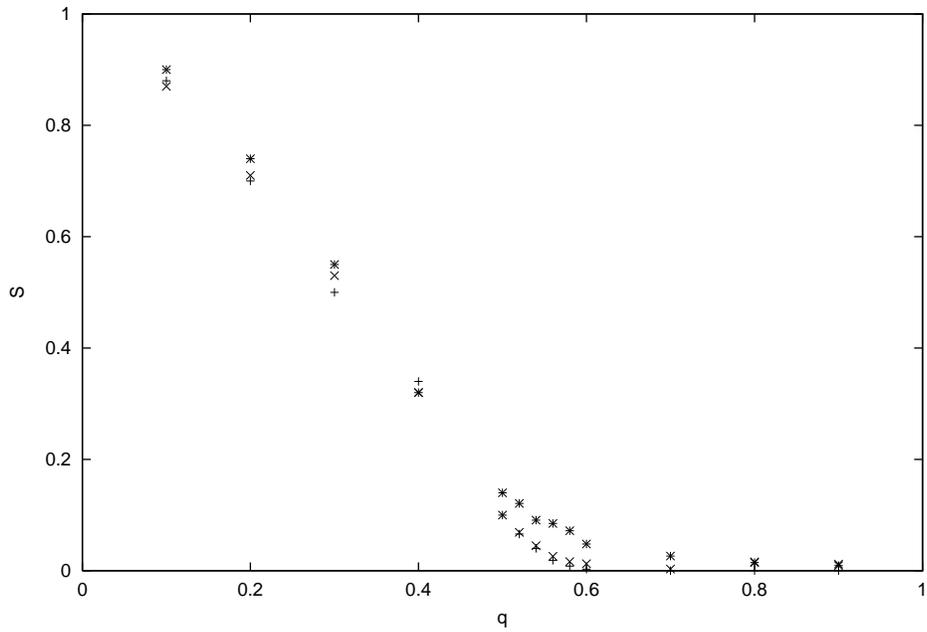}
\end{center}
\caption{Final fraction of visited sites, when a fraction $q$ of the $n$ network
sites is removed; the removed $qN$ sites are the first who joined the network;
$ m=1, \,  \, t \le 10^7, \, P=0.1$. The network size is $N = 10^4 (+), \, 10^3
(\times), \, 10^2$ (stars).  The results for $m=5$ do not differ much.}
\end{figure}

\end{document}